\documentclass[10pt,a4paper]{article}

\date{}
\begin{document}

\title{About the  mass problem\\
{\normalsize Physics and geometry}}
\author{Ll. Bel\thanks{e-mail:  wtpbedil@lg.ehu.es} \and J. Mart\'in--Mart\'{\i}n\thanks{Departamento de F\'{\i}sica
Fundamental, Universidad de Salamanca, 37008 Salamanca, Spain,
e-mail: chmm@usal.es}}

\maketitle

\begin{abstract}

The concept of {\it active gravitational mass}, its definition and its relation with the sources of a gravitational field, was clearly established by Tolman in 1934. On the contrary, and surprisingly in our opinion, the concept of {\it proper mass} has remained obscure in General relativity. We compare a new definition to an apparent more obvious one and discuss how each choice modifies the ratio of active to proper gravitational mass.

\end{abstract}

\section*{Introduction}

This paper deals exclusively with Schwarzschild's complete solution of Einstein's equations, i.e., the solution generated by a spherically symmetric body with uniform mass density $\rho$ and finite radius.

Sections 1 and 2 are a summary of known results. These include Tolman's identification of the {\it active gravitational mass} $M_a$ and the description of the line-element to be used, both in Droste-Hilbert coordinates and harmonic ones, depending on whether technical simplicity or physical meaning is needed.

In Section 3 we propose two definitions of {\it proper gravitational mass}. This is a concept that depends both on the physics that the density $\rho$ describes as well as on the geometry that provides the volume element necessary to the integration over the total volume of the source.

We consider two geometries: the quotient space geometry of the space-time and a geometry conformal to it. It is this second geometry in conjunction with the use of harmonic coordinates  that guarantees a smooth transition from General Relativity to Newtonian theory for weak gravitational fields.

$\hat M_p$ being the proper mass derived from the quotient geometry and $\bar M_p$ the proper mass derived from the conformal geometry we prove in Section 4 that $\Delta\hat M_p=M_a-\hat M_p$ is negative and $\Delta\bar M_p=M_a-\bar M_p$ is positive so that $\hat M_p>M_a>\hat M_p$. We prove also that both are integrals, over all space, derived respectively from negative and positive densities.

In Section 5 we compare again $\hat M_p$ and $\bar M_p$ to $M_a$ both at the linear approximation and numerically for some particular values of the compactness parameter of the source.

In our conclusion we give a hint about how to answer the question: are there any observational effects related to an appropriate choice of the concept of proper gravitational mass?

\section{Active gravitational mass}

We discuss the mass-problem considering only static spherically  symmetric bo\-unded sources with uniform {\it proper mass density}. If we imagine the source as a juxtaposition of small elements, by proper mass we mean here  the sum of the masses of the individual elements measured before being juxtaposed.

The solution to this problem was first obtained by Schwarzschild \cite{Schwarzschild1}\cite{Schwarzschild2} using a rather inconvenient system of coordinates that we shall not use here despite its historical interest and physical insight. We consider to begin with the rather general form of the line-element:

\begin{equation}
\label{1.1}
ds^2=-A^2c^2dt^2+d\hat s^2, \quad A>0
\end{equation}
where, using polar coordinates:

\begin{equation}
\label{1.1B}
d\hat s^2=B^2dr^2+BCr^2(d\theta^2+\sin^2\theta d\phi^2), \quad B,\ C>0
\end{equation}
or equivalently, using Cartesian coordinates:

\begin{equation}
\label{1.1A}
d\hat s^2=B\left[C\delta_{ij}+(B-C)\frac{x_ix_j}{r^2}\right]dx^idx^j,
\ \ i,j=1,2,3
\end{equation}
is the quotient 3-dimensional space metric.

The functions $A,\ B$ and $\ C$ are functions of the radial coordinate $r$ satisfying the following conditions.

\begin{itemize}
\item The derivatives at the origin are zero:

\begin{equation}
\label{1.2}
A^\prime(0)=0, \ B^\prime(0)=0, \ C^\prime(0)=0
\end{equation}

\item Their behavior  at infinity is the following:

\begin{equation}
\label{1.3}
A\approx 1-\frac{GM_a}{rc^2}, \  B\approx 1+\frac{GM_a}{rc^2}, \ C\approx 1-\frac{GM_a}{rc^2},
\end{equation}
where $M_a$ is the {\it active gravitational mass}. It can be defined by:

\begin{equation}
\label{1.3A}
\frac{GM_a}{c^2}=\lim_{r\rightarrow\infty} (A^\prime r^2), \ \hbox{or} \
\ \frac{GM_a}{c^2}=\lim_{r\rightarrow\infty} ((1-B^{-1})r)
\end{equation}

\item The exterior and interior line-elements have to be minimally matched in  the  radius $R$ of the central body. This meaning that they have to satisfy the conditions:

\begin{equation}
\label{1.4}
A_i(R)=A_e(R), \ B_i(R)=B_e(R), \ C_i(R)=C_e(R),
\end{equation}
and:

\begin{equation}
\label{1.4A}
A^\prime_i(R)=A^\prime_e(R), \ C^\prime_i(R)=C^\prime_e(R)
\end{equation}
We remind here that Lichnerowicz's matching conditions require also that:

\begin{equation}
\label{1.4B}
B^\prime_i(R)=B^\prime_e(R)
\end{equation}
This is a condition that is compatible with some global coordinate conditions and it is not with others.
\end{itemize}

Schwarzshild's line--element is a solution of Einstein's equations:

\begin{equation}
\label{1.5}
S^0_0=\kappa\rho, \ \quad S^1_1=S^2_2=S^3_3=-\kappa P(r), \ \ \kappa=\frac{8\pi G}{c^2}
\end{equation}
where $P$ is the pressure inside the source; it is zero outside and on the border.
Then under minimal matching conditions Tolman \cite{Tolman}  proved that:

\begin{equation}
\label{1.6}
M_a=\int_{Vol} (\rho+3P) A\sqrt{\hat g}\,dx^1dx^2dx^3, \ \ \hat g=\det(\hat g_{ij})
\end{equation}
or:

\begin{equation}
\label{1.6A}
M_a=4\pi\int_0^R (\rho+3P) AB^2Cr^2\,dr
\end{equation}

\section{Systems of coordinates}

For our purpose we need to mention two systems of coordinates. The first one is the Droste-Hilbert one \cite{Droste}\cite{Hilbert}, which is characterized by the condition:

\begin{equation}
\label{0.1}
BC=1
\end{equation}
For this system of coordinates the functions $A, B, C$ and $P$ are explicitly known. One has in the interior:

\begin{equation}
\label{0.2}
A_i=\frac32\sqrt{1-qR^2}-\frac12\sqrt{1-qr^2}, \ B_i=(\sqrt{1-qr^2})^{-1},
\ \ q=\frac13\kappa\rho
\end{equation}
and:

\begin{equation}
\label{0.3}
P_i=\frac{\rho}{c^2}\frac{\sqrt{1-qr^2}-\sqrt{1-qR^2}}{3\sqrt{1-qR^2}-\sqrt{1-qr^2}}
\end{equation}
and in the exterior:

\begin{equation}
\label{0.4}
A_e=\sqrt{1-\frac{\kappa M_a}{4\pi r}}, \ B_e=A_e^{-1}
\end{equation}
In this case the exterior and interior expressions are matched only minimally on the border.

The second system of coordinates that we have to mention is that for which the coordinates $x^i$ are harmonic. The condition that implements this requirement is:

\begin{equation}
\label{0.5}
C^\prime=\frac{2}{r}(B-C)-C\frac{A^\prime}{A}
\end{equation}
It has been proved in \cite{Chus} that functions $A,B,C$ and $P$ can be found, both in the interior and exterior, satisfying the requirements of the preceding section, as well as  the above one, that match on the border in the sense of Lichnerowicz.

Considering $A, B, C$ as functions of $\kappa$, and solving Einstein's equations at first order one has:

\begin{equation}
\label{0.6}
A_i=1+\frac{\kappa}{12}\rho(r^2-3R^2)\,, \ B_i=C_i=-A_i+2\,, \ P_i=\frac{\kappa\rho^2}{12c^2}(R^2-r^2)
\end{equation}
and:

\begin{equation}
\label{0.7}
A_e=1-\frac{\kappa}{6}\rho\frac{R^3}{r}, \ B_e=C_e=-A_e+2
\end{equation}

\section{Two definitions of proper mass}

In the preceding section we said what we meant by proper mass, but we did not say how to derive from it how to define the effective density that takes into account the geometry of space inside the source. The only {\it a priori} restriction to be satisfied is that it has to lead to an admissible formula of the following type:

\begin{equation}
\label{2.1}
M_p=\int_{Vol} \rho f\,dx^1dx^2dx^3
\end{equation}
$f$ being some geometrical volume density. From whatever choice me make will depend the relationship between $M_p$ and $M_a$

We discuss below two  judicious choices:

\begin{itemize}
\item $f$ is the density volume derived from the quotient metric, i.e. $f=\sqrt{\hat g}$. So that in polar coordinates we would have

\begin{equation}
\label{2.2}
\hat M_p=4\pi\int_0^R \rho B^2Cr^2\,dr
\end{equation}
This will be considered an obvious choice by all those who take for granted that $d\hat s^2$ describes the real geometry of space inside and outside the source. But we want to remind that no real test whatsoever has up to now sustained this belief. Therefore any other reasonable choice must be considered.

\item In this case $f$ is the volume density associated  with a metric conformal to $d\hat s^2$,

\begin{equation}
\label{2.3}
d\bar s^2= A^2 d\hat s^2
\end{equation}
which leads to the following definition:

\begin{equation}
\label{2.4}
\bar M_p=4\pi\int_0^R \rho A^3B^2Cr^2\,dr
\end{equation}

Ehlers and Kund,\cite{Ehlers}, and later Geroch, \cite{Geroch}, realized that introducing the metric $d\hat s^2$ has some technical advantages, but what it really means from a physical point of view was initially realized by Fock, \cite{Fock}, and developed by one of us, \cite{Bel69}-\cite{Bel09}.
\end{itemize}

It follows from (\ref{0.6}) and (\ref{0.7}) that at the first order we have for $d\hat s^2$
\begin{equation}
\label{2.5}
d\hat s_i^2=\left[1+\frac{\kappa\rho}{6}(3R^2-r^2)\right]\big[dr^2+r^2(d\theta^2+\sin^2\theta\phi^2)\big]
\end{equation}
and:

\begin{equation}
\label{2.5A}
d\hat s_e^2=\left(1+\frac{\kappa\rho}{3}\frac{R^3}{r}\right)\big[dr^2+r^2(d\theta^2+\sin^2\theta\phi^2)\big]
\end{equation}
while for $d\bar s^2$ we have inside and outside:

\begin{equation}
\label{2.6}
d\bar s^2=dr^2+r^2(d\theta^2+\sin^2\theta\phi^2)
\end{equation}
this meaning that choosing $d\bar s^2$ to describe the global geometry of space complements the Poisson equation:

\begin{equation}
\label{2.6.1}
A^{\prime\prime}+\frac{2}{r}A^\prime=4\pi G\rho
\end{equation}
and substantiates the fact that at the first approximation the Schwarzschild model of space-time coincides with the corresponding Newtonian one. This is not true when a different geometry of space or a different system of space coordinates (not harmonic) is used, and should be kept in mind when describing general relativistic effects as deviations from Newtonian gravity.

\section{Active mass versus proper mass}

Using Droste coordinates, let us consider the following general definition of proper mass:

\begin{equation}
\label{Mass0}
M_p=4\pi\int_0^R (\rho r )(F^3Br)\,dr
\end{equation}
so that if $F=1$ then $M_p=\hat M_p$ and if $F=A$ then $M_p=\bar M_p$. The first Einstein equation of (\ref{1.5}) is explicitly:

\begin{equation}
\label{Field equation}
2B^{-3}B^\prime+r^{-1}(1-B^{-2})=\rho r,  \quad 8\pi G =1
\end{equation}
so that for $r<R$ we can write $M_p$ as:

\begin{equation}
\label{Mass1.i}
M_p=4\pi\int_0^R \big[2B^{-3}B^\prime+r^{-1}(1-B^{-2})\big](F^3Br)\,dr
\end{equation}
while for $r>R$ we have trivially:

\begin{equation}
\label{Mass1.e}
0=4\pi\int_R^\infty\big[2B^{-3}B^\prime+r^{-1}(1-B^{-2})\big](F^3Br)\,dr
\end{equation}
Adding (\ref{Mass1.i}) and (\ref{Mass1.e}) we get:

\begin{equation}
\label{A1}
M_p=4\pi\int_0^\infty (2B^{-3}B^\prime+r^{-1}(1-B^{-2}))(F^3Br)\,dr
\end{equation}

Using now:

\begin{equation}
\label{A2}
2F^3(B^{-2}B^\prime)r=-2(F^3B^{-1}r)^\prime+2B^{-1}(F^3r)^\prime
\end{equation}
we obtain:
\begin{equation}
\label{A3}
M_p=4\pi\int_0^\infty (-2(F^3B^{-1}r)^\prime+6F^2B^{-1}F^\prime r+F^3(B+B^{-1}) \,dr
\end{equation}
that can be written also as:

\begin{equation}
\label{A4}
M_p=4\pi\int_0^\infty \Big[2(F^3(1-B^{-1})r)^\prime+6F^2(B^{-1}-1)F^\prime r+F^3(B+B^{-1}-2)\Big] \,dr
\end{equation}
and therefore:

\begin{equation}
\label{Mass3}
M_p=8\pi\lim_{r\rightarrow \infty}\big[F^3(1-B^{-1})r\big]+4\pi\int_0^\infty \big[F^3(B+B^{-1}-2)+6F^2F^\prime(B^{-1}-1)r\big]dr
\end{equation}

Taking into account (\ref{1.3}), if $F=1$ we get, after some re-organization, the formula:

\begin{equation}
\label{Quotient mass}
M_a=\hat M_p+4\pi\int_0^\infty \hat\rho_g r^2\,dr\,, \quad \hat\rho_g=-\frac{1}{Br^2}(B+B^{-1}-2)
\end{equation}
while if $F=A$ we get:

\begin{equation}
\label{Conformal  mass}
M_a=\bar M_p+4\pi\int_0^\infty \bar\rho_g r^2\,dr, \quad \bar\rho_g=\frac{1}{A^3Br^2}
(6A^2A^\prime(1-B^{-1})r-A^3(B+B^{-1}-2)
\end{equation}

Since $B$ is positive and:

\begin{equation}
\label{Quotient  mass 2}
B+B^{-1}-2=B(1-B^{-1})^2>0
\end{equation}
$\hat\rho_g$ is definite negative.
On the contrary $\bar\rho_g$ is definite positive. In fact we may write:

\begin{equation}
\label{Conformal  mass 2}
\bar\rho_g=\frac{1}{r^2}(1-B^{-1})^2(Z-1)
\end{equation}
where :

\begin{equation}
\label{Z}
Z=\frac{6A^{\prime}r}{A(B-1)}
\end{equation}
and therefore $\bar\rho_g$ will be positive for those values of $r$ where $Z$ is greater than $1$. From (\ref{0.2}) and (\ref{0.4}) it can be checked easily  that for all values of r we have:

\begin{equation}
\label{Aprime}
A^{\prime}=\frac{B^2-1}{2Br}
\end{equation}
so that when $r<R$ we have:

\begin{equation}
\label{Interior}
Z_i=\frac{3(B_i+1)}{A_iB_i}
\end{equation}
Since:

\begin{equation}
\label{AB}
A_iB_i=\frac{3\sqrt{1-qR^2}}{2\sqrt{1-qr^2}}-\frac12
\end{equation}
the maximum value of $A_iB_i$ is $1$, reached at $r=R$, and the minimum value of $B_i$ is $1$, reached for $r=0$. Therefore $Z_i$ is always greater than $3$

For $r>R$ we have:

\begin{equation}
\label{Exterior}
Z_e=3(B_e+1)
\end{equation}
and since the minimum asymptotic value of $B_e$ is $1$, $Z_e$ is always greater than $6$. This proves our assertion.

It follows from these results that identifying $M_p$ with $\hat M_p$  suggests that the work done by the gravitational field in a collapse process, otherwise called the potential energy of the source, is reflected by a localized loss of mass. On the other hand identifying $M_p$ with $\bar M_p$suggests that the mass-equivalent gravitational energy could contribute to enhance the strength of the gravitational field of a bare source, as it was also suggested in \cite{Bel09} for Newtonian gravity. Another possible suggestion is that we should consider $\hat M_p$ and $\bar M_p$ as independent physical concepts.

\section{Numerical estimates}

We proceed now to discuss the ratio $M_a/M_p$ using Tolman's formula (\ref{1.6}) together with the definitions (\ref{2.2}) and (\ref{2.4}).

For weak gravitational fields we can use (\ref{0.6}). We obtain then at the first order of approximation:

\begin{eqnarray}
\label{2.8Da}
\hat X&\equiv& \frac{M_a}{\hat M_p}=1-\frac{1}{10}\kappa\rho R^2\\
\label{2.8Db}
\bar X&\equiv& \frac{M_a}{\bar M_p}=1+\kappa\frac12\rho R^2
\end{eqnarray}

To deal with stronger gravitational fields it is better to rely on Droste-Hilbert coordinates. As it is well-known,  Eq. (\ref{1.6}) is equivalent to

\begin{equation}
\label{2.8A}
M_a=4\pi\int_0^R \rho r^2dr=\frac43\pi\rho R^3
\end{equation}
and the two definitions of $M_p$ become respectively:

\begin{equation}
\label{2.7}
\hat M_p=4\pi\int_0^R \rho B_ir^2dr
\end{equation}
and:

\begin{equation}
\label{2.8}
\bar M_p=4\pi\int_0^R \rho A_i^3B_ir^2dr
\end{equation}
Let us consider the compactness parameter:

\begin{equation}
\label{2.9}
\lambda=\frac{2GM_a}{Rc^2}
\end{equation}
Using units such that:

\begin{equation}
\label{2.9A}
8\pi G=1, \ c=1, \ M_a=4\pi
\end{equation}
we have:

\begin{equation}
\label{2.9B}
R=\lambda^{-1}, \ \rho=3\lambda^3
\end{equation}
and we can proceed to calculate numerically  the integrals  defining $\hat M_p$ and $\bar M_p$ giving to $\lambda$ desired values. The following are some of the results that we have obtained:

\vspace{.5cm}

\begin{tabular}{|l|l|l|}
\hline
\\[-2ex]
$\quad\lambda$ & $\quad\hat X$ & $\quad\bar X$
\\[.5ex]
\hline
8/9    & \ .60948 & \ 51.7311\\
1/2    & \ .82587 & \ 3.00745\\
1.0\,e-1 & \ .96926 & \ 1.17277\\
1.0\,e-2 & \ .99699 & \ 1.01520\\
1.0\,e-3 & \ .99970 & \ 1.00150\\
1.0\,e-4 & \ .99997 & \ 1.00013\\

\hline
\end{tabular}

\vspace{.5cm}

$\hat X$ and $\bar X$ are dimensionless quantities, therefore the values listed above are independent of the system of units mentioned in (\ref{2.9A}). They are also valid  for any of the system of coordinates considered in Section 1.

The geometrical considerations discussed in this paper can not settle the problem of deciding what is the correct choice to define $M_p$, but we think that it would be reckless to ignore them while we wait for a happy discovery that validates one of them.

\section{Concluding remarks}

At the beginning of this paper we gave an operational definition of gravitational proper mass that required to divide the source of the gravitational field into pieces to be weighted one by one. This can be done only under appropriate laboratory conditions in which case the gravitational field will be so weak that it will be illusory to distinguish proper mass from active gravitational mass.

At the astrophysical or cosmological level on the other hand while our operational definition is useless there is an urgent need to be able to compare the active gravitational mass to the proper mass derived from the {\it visible mass}. This invites us to ask the following question: are there observable effects that could favor one or the other of the two definitions of proper mass that we have proposed?  We offer one possibility for those cases where the visible mass can reliably suggest the amount of proper mass involved. Namely to accept tentatively that if $\rho$ is the proper mass density the solution of Einstein's equations to be used to describe the gravitational field is not that generated by $\rho$ but that generated by an effective mass density $\rho+\rho_g$ with $\rho_g$ equal to $\hat\rho_g$ or $\bar\rho_g$ as defined in Section 4. This essentially does not change General Relativity but it changes the conventional relationship between the sources and the gravitational field that they generate.

As illustrate  by the table of values of the preceding Section the active gravitational field $M_a$ remains very close to $\hat M_p$ but may differ very much from $\bar M_p$ for very extreme objects with compactness parameter close to 1, and it might be safe to keep in mind that including in $\rho$ the internal energy derived from the physical processes going on in the source, the class of relevant interesting objects can be larger than otherwise expected. If this is the case, choosing a good definition of proper mass may become a crucial problem.

\section*{Appendix}

We use in this appendix the notations $r_d$ and $R_d$ to refer to the radial variable of Droste's coordinates and to the value of $r_d$ at the frontier of the source of Schwarzschild's solution. We use the same sub-index to refer to those quantities that are functions of $r_d$ and $R_d$. The corresponding quantities, $r$, $R$, and functions of, referred to the system of polar global harmonic coordinates, whose existence was proved in \cite{Chus}, are written without sub-index.

The relationship between the pair ($r_d, \ R_d$) and the pair ($r, \ R$) defines a coordinate transformation:

\begin{equation}
\label{Bel1}
r_d=r_d(r,R), \ \ R_d=R_d(R)
\end{equation}
that induces the following transformations of the metric coefficients:

\begin{eqnarray}
\label{Bel2}
A(r,R)&=&A_d[r_d(r,R),R_d(R)], \\
B(r,R)&=&B_d[r_d(r,R),R_d(R)]\frac{dr_d(r,R)}{dr}, \\
C(r,R)&=&\frac{r_d^2(r,R)}{r^2B(r,R)}
\end{eqnarray}
Any other scalar under such a coordinate transformation, like $A$, will transform accordingly.

Using a system of units such that $8\pi G=c=1$, the following formulas are correct up to the third order in powers of the density $\rho$:

\begin{itemize}

\item $r<R\,$:

\begin{eqnarray}
\label{Ap1}
&&\hspace{-1.2cm}r_d = r\left[1+\frac14\left(-\frac13r^2 + R^2 \right)\rho +\frac{1}{12}\left(\frac1{14}r^4 - \frac7{10}r^2R^2 + \frac53R^4 \right)\rho^2 \right.\nonumber
\\[1.5ex]
&&\hspace{-0.5cm}\left.+\frac{1}{9}\left(-\frac{11}{6048}r^6 + \frac{17}{320}r^4R^2 -\frac25 r^2R^4 +\frac{17963}{20160} R^6 \right)\rho^3+ O(\rho^4) \right]
\end{eqnarray}

\item $r>R\,$:
\begin{eqnarray}
\label{Ap1}
&&
\hspace{-2cm}r_d = r\left[1+\frac16\frac{R^3}{r^3}\rho +\left(\frac1{12}\frac{R^5}{r} + \frac1{315}\frac{R^7}{r^3} \right)\rho^2
\right.\nonumber
\\[1.5ex]
&&\hspace{1cm}\left.
+\left(\frac{2}{35}\frac{R^7}{r} + \frac{53}{17010}\frac{R^9}{r^3}\right)\rho^3+ O(\rho^4) \right]
\end{eqnarray}
\end{itemize}
\medskip
where in both cases we have:
\begin{equation}
R_d = R\left[1+\frac16 R^2\rho + \frac{109}{1260}R^4\rho^2 + \frac{205}{3402}R^6\rho^3 +  O(\rho^4) \right]
\end{equation}

The corresponding metric functions $A,B$ and $C$ are .
\begin{itemize}

\item $r<R\,$:
\begin{eqnarray}
\label{Ap2a}
&&\hspace{-1cm}A=1+\frac{1}{12}(\eta^2-3)\rho R^2 -\frac{1}{144}
(\eta^4-6\eta^2+15) \rho^2 R^4 \nonumber
\\[1ex]
&&\ + \frac{1}{432}\left(\frac5{28}\eta^6  -\frac{27}{10}\eta^4 + \frac{49}4\eta^2 - \frac{2043}{70}\right)   \rho^3 R^6 + O(\rho^4)
\\[2ex]
\label{Ap2b}
&&\hspace{-1cm}B=1-\frac{1}{12}(\eta^2-3)\rho R^2 +\frac{1}{36}
\left(\frac1{14}\eta^4-\frac95\eta^2+5\right) \rho^2 R^4  \nonumber
\\[1ex]
&&\ + \frac{1}{864}\left(\frac79\eta^6  -\frac32\eta^4 - \frac{141}5\eta^2 + \frac{17963}{210}\right)   \rho^3 R^6 + O(\rho^4)
\\[2ex]
\label{Ap2c}
&&\hspace{-1cm}C=1-\frac{1}{12}(\eta^2-3)\rho R^2 +\frac{1}{36}
\left(\frac5{14}\eta^4-\frac{12}5\eta^2+5\right) \rho^2 R^4  \nonumber
\\[1ex]
&&\ + \frac{1}{32}\left(\frac{71}{1701}\eta^6  -\frac{13}{30}\eta^4 + \frac95\eta^2 - \frac{17963}{5670}\right)  \rho^3 R^6 + O(\rho^4)
\end{eqnarray}

\item $r>R\,$:

\begin{eqnarray}
\label{Ap3a}
&&\hspace{-1cm}A=1-\frac{1}{6\eta}\rho R^2 +\frac{1}{36}
\left(\frac1{2\eta^2} -\frac3{\eta}\right)\rho^2 R^4 \nonumber
\\[1ex]
&&\ + \frac{1}{216}\left(\frac4{35\eta^4}  -\frac1{2\eta^3} +\frac3{\eta^2}- \frac{432}{35\eta}\right)  \rho^3 R^6 + O(\rho^4)
\\[2ex]
\label{Ap3b}
&&\hspace{-1cm}B=1+\frac{1}{6\eta}\rho R^2 +\frac{1}{36}
\left(-\frac8{35\eta^3} +\frac1{2\eta^2}+\frac3{\eta}\right) \rho^2 R^4 \nonumber
\\[1ex]
&&\ + \frac{1}{216}\left(-\frac{12}{35\eta^4}  -\frac{533}{630\eta^3} +\frac3{\eta^2}+ \frac{432}{35\eta}\right)   \rho^3 R^6 + O(\rho^4)
\\[2ex]
\label{Ap3c}
&&\hspace{-1cm}C=1+\frac{1}{6\eta}\rho R^2 +\frac{1}{36}
\left(\frac{16}{35\eta^3} -\frac1{2\eta^2} + \frac3{\eta}\right) \rho^2 R^4 \nonumber
\\[1ex]
&&\ + \frac{1}{216}\left(\frac{12}{35\eta^4}  + \frac{1381}{630\eta^3} -\frac3{\eta^2}+ \frac{432}{35\eta}\right)   \rho^3 R^6 + O(\rho^4)
\end{eqnarray}

\end{itemize}
 where we have used the dimensionless variable $\eta\equiv r/R$ whose value is $1$ at the frontier of the source.

The remaining relevant quantities introduced in the main body of the text are at the same order of approximation:

\begin{itemize}

\item $r<R\,$:

\begin{eqnarray}
&&\hspace{-1.2cm}M_a =\frac43\pi \rho R_d^3 \nonumber
\\[1ex]
&&\hspace{-0.6cm}= \frac43 \pi\rho R^3\left[1 + \frac12 R^2\rho +
\frac{12}{35} R^4\rho^2 +
\frac{881}{3240} R^6\rho^3+ O(\rho^4)\right]
\end{eqnarray}

\end{itemize}

\begin{eqnarray}
\label{Ap2a}
&&\hspace{-1cm}\hat X\equiv \frac{M_a}{\hat M_p}=1-\frac{1}{10} R^2\rho -\frac{173}{4200} R^4\rho^2 - \frac{30169}{1134000}R^6\rho^3  + O(\rho^4)
\\[1.5ex]
\label{Ap2b}
&&\hspace{-1cm}\bar X\equiv \frac{M_a}{\bar M_p}=1+\frac12R^2\rho  +\frac{163}{420} R^4\rho^2 + \frac{143}{420}R^6 \rho^3 + O(\rho^4)
\end{eqnarray}
\begin{itemize}

\item $r<R\,$:

\begin{eqnarray}
&&\hspace{-1cm}\hat\rho_g = -\frac1{36}r^2\rho^2 -\frac1{72}r^2R^2\rho^3+ O(\rho^4)
\\[1ex]
&&\hspace{-1cm}\bar\rho_g = \frac5{36}r^2\rho^2 +\frac1{36}r^2(-r^2+4R^2)\rho^3+ O(\rho^4)
\end{eqnarray}

\item $r>R\,$:

\begin{eqnarray}
&&\hat\rho_g = -\frac1{36}\frac{R^6}{r^4}\rho^2 -R^8\left(\frac1{36r^4}-\frac{R}{72r^5}\right)\rho^3+ O(\rho^4)
\\[1ex]
&&\bar\rho_g = \frac5{36}\frac{R^6}{r^4}\rho^2 +R^8\left(\frac5{36r^4}-\frac{R}{18r^5}\right)\rho^3+ O(\rho^4)
\end{eqnarray}
\end{itemize}

\section*{Acknowledgements}

One of us (J. Mart\'{\i}n) acknowledge financial support under the projects
FIS2006-05319 of the Spanish MEC and SA010CO5 of the Junta de Castilla y
Le\'on.

\end{document}